# Chirality-dependent spin current generation in a helimagnet: zero-field probe of chirality

[1]Institute for Materials Research, Tohoku University, Sendai, Japan.
[2]Department of Physics, Toho University, Funabashi, Japan.
[3]PRESTO, Japan Science and Technology Agency, Kawaguchi, Japan.
[4]Center for Science and Innovation in Spintronics, Tohoku University, Sendai, Japan.
[5]Advanced Science Research Center, Japan Atomic Energy Agency, Tokai, Japan.

Hidetoshi Masuda[1*], Takeshi Seki[1*], Jun-ichiro Ohe[2], Yoichi Nii[1,3], Koki Takanashi[1,4,5], Yoshinori Onose[1*]

**In a magnetic texture, the spin of a conduction electron is forced to be aligned to the localized moment. As a result, the topology of the magnetic texture affects the electron dynamics in nontrivial ways. A representative example is the topological Hall effect in noncoplanar spin textures with finite spin chirality. While propagating in the noncoplanar spin texture, electrons acquire Berry phase, and their motion is deflected as if they were in a magnetic field. Here, we report a distinct Berry phase effect in a coplanar helimagnet: the spin moment of the conduction electron is polarized under electric currents depending on the chirality of the helimagnet. The accumulated spin polarization works as a source of spin current, and the chirality can be detected by the inverse spin Hall mechanism. The functionality allows us to read out the chirality without magnetic fields, and therefore paves the way to future helimagnet-based spintronics.**

The detection of magnetic states is one of the central issues in spintronics. The giant magnetoresistance in multilayers composed of alternating ferromagnetic and non-magnetic conductive layers has been commercially applied to magnetic sensors that can effectively read out the ferromagnetic domain information stored in hard disk drives[1,2]. The much larger magnetoresistance in a magnetic tunnel junction, called tunnel magnetoresistance, is utilized for reading the stored information in magnetoresistive random access memory (MRAM)[3]. In recent years, spintronics based on antiferromagnets has been attracting considerable attention to overcome drawbacks of ferromagnet-based spintronics, such as the stray field problem and the low switching speed[4-6]. As a sensitive probe of antiferromagnetic states, researchers have extensively investigated responses depending on the antiferromagnetic states, as exemplified by anisotropic magnetoresistance[7-10], spin Hall magnetoresistance[11-13], and the anomalous Hall effect[14-16]. Quite recently, we succeeded in controlling the chirality of a metallic helimagnet, MnP, around 50 K[17], and subsequently, the functionality was demonstrated for the thin film of another helimagnet, MnAu$_2$, even at room temperature[18]. A helimagnet is one form of antiferromagnet that hosts the chirality (or helicity) degree of freedom corresponding to the right- or left-handedness of the spiral rotation of the localized moments[19]. Because the chirality degree of freedom works as a stable information carrier robust against external perturbation, these studies may open up the possibility of helimagnet-based spintronics. However, the chirality was previously probed by the field-asymmetric component of second harmonic resistivity (nonreciprocal electric transport), which cannot be directly applied to commercial devices. In this paper, we demonstrate that the chirality can be detected by a Hall-like signal in a bilayer device composed of the helimagnetic MnAu$_2$ and a spin Hall material Pt.

The origin of the Hall-like signal can be explained by the following two steps. First, an electric current along the helimagnetic wave vector induces spin accumulation depending on the chirality in the helimagnetic MnAu$_2$ layer. Next, the accumulated spin is diffused to the Pt layer and induces a transverse voltage by means of the inverse spin Hall effect[20-22]. While the mechanism of the second step is well known[23,24], we explain the first step microscopically below.

We schematically depict the phenomenon of current-induced spin polarization in Fig. 1a. The spin of the conduction electrons is aligned to the localized moments owing to the *s-d* exchange coupling $J_{sd}$. When a conduction electron propagates along the propagation vector of the helimagnet, the spin rotates around the propagation vector with a sense that depends on the chirality $\lambda$. The chirality-dependent spin rotation affects the dynamics of the conduction electron through the fictitious vector potential (Berry connection) $\vec{A} = (i\lambda\hat{\sigma}_x, 0, 0)$, where the *x*-axis is along the propagation vector of the helimagnet and $\hat{\sigma}_x$ is the Pauli matrix[25,26]. Therefore, when an electric current $\vec{j} = (j_x, 0, 0)$ is applied, the current dependent energy $\Delta E \sim <\vec{p}\cdot\vec{A}> \sim \vec{j}\cdot\vec{A} \sim \lambda\hat{\sigma}_x j_x$ gives rise to spin polarization depending on the chirality. In the literature[25], this phenomenology was theoretically discussed for a helimagnet induced by the Dzyaloshinskii–Moriya interaction, in which the chirality is fixed. To confirm the validity for a variable-chirality system, we numerically calculated the electric current-induced spin polarization in a classical helimagnetic model with a system size of

100×30 sites, in which the nearest (next-nearest) neighbor magnetic interaction $J_1$ ($J_2$) of localized moments is ferromagnetic (antiferromagnetic), and $J_{sd}$ couples the spin of the conduction electron and the localized moment. The details of the numerical calculation are shown in the Supplementary Information. Figure 1b shows the calculated chirality $\lambda = \left(\vec{S} \times \frac{\partial \vec{S}}{\partial x}\right)_x$ and spin polarization of conductance $P_x = \text{Tr}(\hat{t}^+ \hat{\sigma}_x \hat{t})/\text{Tr}(\hat{t}^+ \hat{t})$ as a function of $|J_2/J_1|$ ($\hat{t}$ is the transmission matrix represented in the 2×2 spin space and $\hat{\sigma}_x$ is the $x$ component of Pauli matrix). $P_x$ measures the spin-polarization of electric current. Therefore, spin accumulation is induced under an electric current when $P_x$ is finite. Chirality $\lambda$ emerges around $|J_2/J_1|$ = 0.05 and increases rapidly around $|J_2/J_1|$ = 0.25. Whereas the helimagnetic state is known to be stabilized for $|J_2/J_1|$ > 0.25 in an infinite $J_1$-$J_2$ classical model[27], the ferromagnetic-helimagnetic transition seems to be broadened in this finite model. The chirality could be controlled by the suitable choice of initial states (for details, see the Supplementary Information). Importantly, $P_x$ that depends on the chirality begins to evolve around $|J_2/J_1|$ = 0.25. Note that the spin polarization oscillation with the ratio $|J_2/J_1|$ seems to be caused by a resonance condition between the Fermi wavelength and the helical pitch. The calculation numerically demonstrates the electric current-induced spin polarization with a sign that depends on the chirality.

Next, let us discuss the experimental results. Figure 1c illustrates the sample device used in this study. We prepared bilayer devices consisting of $MnAu_2$ and Pt on hexagonal $ScMgAlO_4$ substrates (see Methods and Extended Data Figure 1). $MnAu_2$ is a metallic helimagnet having a centrosymmetric tetragonal crystal structure with the space group *I4/mmm*[28-30]. The Mn moments show a helical magnetic order with the propagation vector along the [001] direction and the helical plane parallel to the (001) plane[29] below the transition temperature $T_c$ = 335 K for the thin film case[18] (see Extended Data Figs. 2a and 2b, $T_c$ = 365 K for the bulk case[29,30]). In the device, the helical propagation vector along the [001] direction is parallel to the film plane. The Pt layer was deposited on $MnAu_2$ to a thickness of 10 nm. When an electric current $I$ is applied along the $MnAu_2$ [001] direction, the chirality-dependent accumulated spin moment should diffuse to the Pt layer, and a transverse voltage $V_T$ is expected to emerge owing to the inverse spin Hall effect (ISHE). Therefore, the chirality should be probed by the transverse resistance $R_T = V_T / I$.

Figure 2 demonstrates the observation of chirality-dependent transverse resistance at

260 K. Previously, we have demonstrated that the chirality can be controlled by the application of a magnetic field and electric current along the helimagnetic propagation vector, depending on whether they are parallel or antiparallel[17,18,31]. The controlled chirality was probed by the nonreciprocal electronic transport (NET)[32], which is the field-asymmetric second harmonic resistance[33]. We first reproduced the previous results for the present sample. For the chirality control, a magnetic field ($H_0$ = +5 T or −5 T) and a dc electric current ($I_0$ = +8 mA or −8 mA) were first applied parallel to the MnAu$_2$ [001] direction. Then, the magnetic field was swept to 0 T, traversing the ferromagnetic-to-helimagnetic transition field $H_c$ = 1.9 T (see Extended Data Figs. 2c and 2d), and finally the current was turned off. To obtain the NET after the chirality control, we measured second the harmonic resistance $R^{2\omega}(H)$ while increasing $H$ from 0 to 5 T after the chirality control, and also measured $R^{2\omega}(H)$ while decreasing $H$ from 0 to −5 T after the separate chirality control, and finally deduced the asymmetric component as $R^{2\omega}_{asym}$ = ($R^{2\omega}(+H) - R^{2\omega}(-H)$)/2. Figures 2a and b show the magnetic field $H$ dependence of $R^{2\omega}_{asym}$ at 260 K (the control magnetic field $H_0 > 0$ for Fig. 2a and $H_0 < 0$ for Fig. 2b). Finite NET signals were clearly observed in the helimagnetic state $|H| < H_c$, and their signs depended on whether the control field $H_0$ and control current $I_0$ are parallel or antiparallel, confirming that the chirality control was successfully reproduced for this sample. In Fig. 2g we plot the averaged NET of $\Delta R^{2\omega}_{asym}(|I_0|) = (R^{2\omega}_{asym}(+I_0) - R^{2\omega}_{asym}(-I_0))/2$ as a function of $|I_0|$. The magnitude of $\Delta R^{2\omega}_{asym}$ monotonically increases with increasing $|I_0|$ and tends to saturate, similarly to the previous result[18].

Having reproduced the chirality control in the present MnAu$_2$ / Pt sample, we now discuss the chirality-dependent transverse resistance in MnAu$_2$. Figures 2c and 2d show the $H$ dependence of the transverse resistance $R_T$ after the chirality control procedure with the magnetic field $H_0$ and the electric current $I_0$ at 260 K. Note that the positive field region and the negative field region were separately measured by sweeping the magnetic field from 0 T just after the chirality control procedure. The main contribution to $R_T$ seems to be from trivial effects such as longitudinal resistance, the Hall effect, and planar Hall effect arising from the misalignments of voltage electrodes and the magnetic field.

The slight difference between the positive and negative $H_0$ (Fig. 2c and Fig. 2d, respectively) for the same chirality should be ascribed to the magnetic hysteresis. Nevertheless, the apparent difference between the positive and negative $I_0$ for the same $H_0$ cannot be ascribed to any trivial effects. Because it is reversed by the reversal of $H_0$, the nontrivial component seems to depend on whether $H_0$ and $I_0$ are parallel or antiparallel, as indicated by the dotted lines in Figs. 2c and 2d. To extract the nontrivial component, we calculate the difference between $R_T$ for positive and negative $I_0$ divided by 2, that is, $\Delta R_T = (R_T(+I_0) - R_T(-I_0))/2$ (Fig. 2e ($H_0 > 0$) and Fig. 2f ($H_0 < 0$)). $\Delta R_T$ shows a maximum around 0 T and decreases with increasing $H$ or decreasing $H$ from $H = 0$ T. Then, it almost vanishes in the induced-ferromagnetic phase. The sign is reversed by the reversal of $H_0$. These properties indicate that $\Delta R_T$ reflects the helimagnetic chirality. To confirm this, we have investigated the $|I_0|$ dependence of $\Delta R_T$ (Fig. 2h). It monotonically increases and tends to saturate around $|I_0| = 6$ mA, similarly to the case of $\Delta R^{2\omega}{}_{asym}$ (Fig. 2g). In fact, $\Delta R_T$ is proportional to $\Delta R^{2\omega}{}_{asym}$, as shown in Fig. 2i. That is to say, $\Delta R_T$ certainly reflects the helimagnetic chirality, which should be caused by the aforementioned spin accumulation mechanism.

Figures 3a-f show the magnetic field dependence of $\Delta R^{2\omega}{}_{asym}$ at various temperatures. For these measurements, the chirality control procedure and the measurements were performed at the same temperature. As reported previously, $\Delta R^{2\omega}{}_{asym}$ is observed in the helimagnetic state below $T_c = 335$ K. In the present work, we have newly found a sign change just below the transition temperature: the sign of $\Delta R^{2\omega}{}_{asym}$ at 330K is opposite to that at 300 K (see also, Extended Data Figure 3). The magnitude shows a maximum around 300 K and decreases toward low temperatures. Finally, it almost vanishes below around 100 K. As shown in Fig. 3g, the temperature dependence of $\Delta R^{2\omega}{}_{asym}$ at 0.5 T clearly shows the sign reversal, a maximum around 300 K, and a gradual decrease toward low temperature. Figures 3h-m show the magnetic field dependences of $\Delta R_T$ at various temperatures. The chirality control procedure and the measurements were performed also at the same temperature. While $\Delta R_T$ is negligible at 340 K, it gradually evolves below the helimagnetic transition temperature. The magnitude of $\Delta R_T$ shows a maximum around $H = 0$ T and is negligible above $H_c$ for all the observed temperatures. Similarly to the $\Delta R^{2\omega}{}_{asym}$ case, a sign reversal just below the transition temperature is observed and the magnitude almost vanishes below around 100 K. We show the temperature dependence of $\Delta R_T$ at 0 T in Fig. 3n. The temperature dependence of $\Delta R_T$ is surprisingly similar to that of $\Delta R^{2\omega}{}_{asym}$; both show sign changes, the magnitudes show a maximum at 260–300 K, and both vanish around 100 K. The similarity implies that the mechanisms of these two phenomena are microscopically related, although the details remain to be clarified. Another important point is that $\Delta R_T$ is observed in a wide temperature range including room temperature, similarly to the $\Delta R^{2\omega}{}_{asym}$ case. Therefore, $\Delta R_T$ is useful as a chirality probe even at room temperature and does not require any magnetic field.

In summary, we have observed chirality-dependent transverse resistance in an $MnAu_2$ / Pt bilayer device. This result demonstrates current-induced spin accumulation originating from the Berry phase effect in the coplanar helimagnetic state, which is distinct from the well-known Berry phase effect of the emergent magnetic field in noncoplanar magnetic textures such as a skyrmion lattice[34-36]. A similar spin-polarization phenomenon known as chirality-induced spin selectivity (CISS) has been observed in chiral crystals and molecules[37,38]. The present observation can be viewed as a full magnetic version of CISS. In a helimagnet with a nonchiral crystal structure, the magnetic CISS enables us to read out the chiral information even in the absence of magnetic fields. Because the transverse voltage is induced at the interface, the magnitude of transverse voltage is expected to become larger as the thickness of the thin film is reduced. If a high-quality ultrathin film can be fabricated, the magnitude might satisfy the requirements for practical application as a probe of chirality-based magnetic memory. In this sense, the present result will pave the way to helimagnet-based spintronics.

**Methods**

**Sample fabrication**

An epitaxial film of $MnAu_2$ with a thickness of 100 nm was deposited on a $ScMgAlO_4$ (10−10) substrate by magnetron sputtering from Mn and Au targets at 400 °C. The film was annealed at 600 °C for 1 hour. A Pt layer was then deposited on the $MnAu_2$ at room temperature. As shown in Extended Data Fig. 1, the X-ray diffraction (XRD) results indicate the epitaxial growth of $MnAu_2$ (110) on the $ScMgAlO_4$ (10−10) substrate, where the $MnAu_2$ [001] and the $ScMgAlO_4$ [0001] directions are parallel to each other. The Pt layer was revealed to be parallel to the (111) plane.

**Longitudinal, transverse, and second-harmonic resistance measurement**

For the transport measurements, the bilayer film samples were patterned into Hall bar devices by a standard photolithography technique and Ar plasma etching. The direction of the electric current was parallel to the $MnAu_2$ [001] direction. The width of the Hall bars was 10 μm, and the distance between two voltage electrodes for the resistivity measurement was 25 μm. Electrical contacts were made by photolithography and electron beam evaporation of Ti (10 nm) / Au (100 nm). The transport measurements were performed in a superconducting magnet. The longitudinal, transverse, and 2nd-harmonic longitudinal ac resistances were measured by utilizing the standard lock-in technique with an electric current frequency of 11.15 Hz.

**Magnetization measurement**

The magnetization measurements were performed using a Magnetic Property Measurement System (Quantum Design).

**Numerical calculation**

The stable magnetic structure in a classical $J_1$-$J_2$ Heisenberg model with a system size of 100×30 sites was numerically calculated by using the Landau-Lifshitz-Gilbert equation and 4th-order Runge-Kutta method. Then, the spin polarization of conductance was calculated by using Green's function method, considering the interaction between the spin conduction electron and the localized moment. The details are shown in the supplemental material.

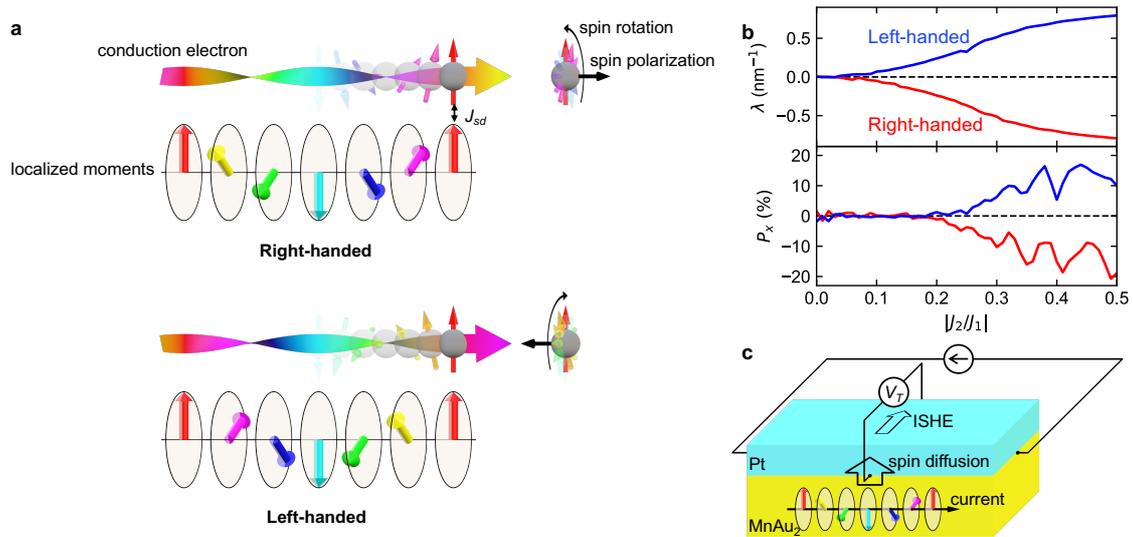

**Figure 1**

**Concept of chirality-dependent transverse resistance.**

**a** Schematic illustration of the spin polarization of an electron moving through a helimagnetic structure.

**b** Calculated chirality $\lambda = \left(\vec{S} \times \frac{\partial \vec{S}}{\partial x}\right)_x$ and spin polarization of conductance $P_x$ along the helimagnetic propagation vector as a function of $|J_2/J_1|$, where $J_1$ ($J_2$) is the ferromagnetic nearest-neighbor (antiferromagnetic next-nearest-neighbor) magnetic interaction of localized moments.

**c** Schematic illustration of the measurement setup for the transverse resistance in the helimagnet $MnAu_2$ / Pt bilayer device. The substrate is not shown for clarity. Electric current is applied parallel to the helimagnetic propagation vector ($MnAu_2$ [001] direction). The electric current in the helimagnetic $MnAu_2$ layer induces a chirality-dependent spin polarization. The accumulated spin polarization diffuses into the Pt layer, inducing the transverse voltage $V_T$ by means of the inverse spin Hall effect (ISHE).

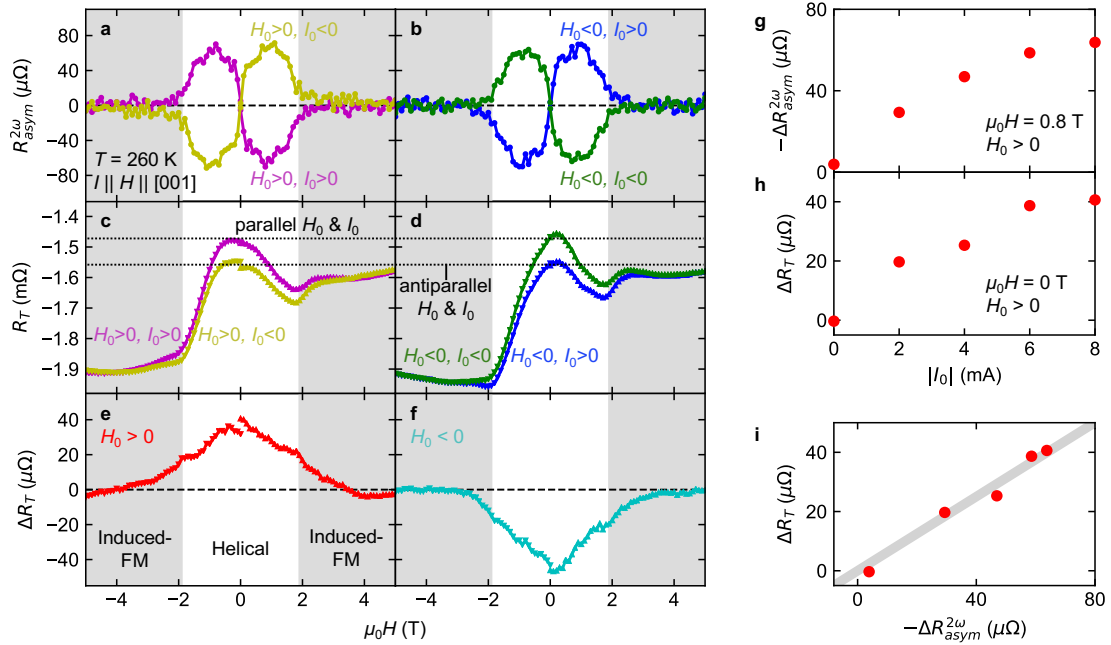

**Figure 2**

**Observation of the chirality-dependent transverse resistance at 260 K.**

**a, b** Magnetic field $H$ dependence of the nonreciprocal electronic transport (NET) $R^{2\omega}_{asym}$ after the chirality control procedure with the magnetic field $H_0$ and electric current $I_0$. The results for $H_0 > 0$ and $H_0 < 0$ are shown in **a** and **b**, respectively. $R^{2\omega}_{asym}$ is the field-antisymmetric component of the second-harmonic longitudinal resistance $(R^{2\omega}(+H) - R^{2\omega}(-H))/2$. While the data of $H < 0$ are merely copies of the $H > 0$ data, we plot the $H < 0$ data just for clarity. The gray shading represents the induced-ferromagnetic (FM) phase.

**c, d** Magnetic field dependence of the transverse resistance $R_T$ after the chirality control with $H_0$ and $I_0$. The results for $H_0 > 0$ and $H_0 < 0$ are shown in **c** and **d**, respectively. The horizontal dotted lines indicate the maxima of $R_T$ for the parallel and antiparallel $H_0$ and $I_0$.

**e, f** Magnetic field dependence of the chirality-dependent component of the transverse resistance estimated from the relation $\Delta R_T = (R_T(+I_0) - R_T(-I_0))/2$ for $H_0 > 0$ and $H_0 < 0$ (**e** and **f**, respectively).

**g** $|I_0|$ dependence of $-\Delta R^{2\omega}_{asym}$ at 0.8 T. Here, $\Delta R^{2\omega}_{asym}$ is the averaged NET signal $(R^{2\omega}_{asym}(+I_0) - R^{2\omega}_{asym}(-I_0))/2$.

**h** $|I_0|$ dependence of $\Delta R_T$ at 0 T.

**i** $\Delta R_T$ at 0 T is plotted against $-\Delta R^{2\omega}_{asym}$ at 0.8 T. The gray line indicates the linear relation.

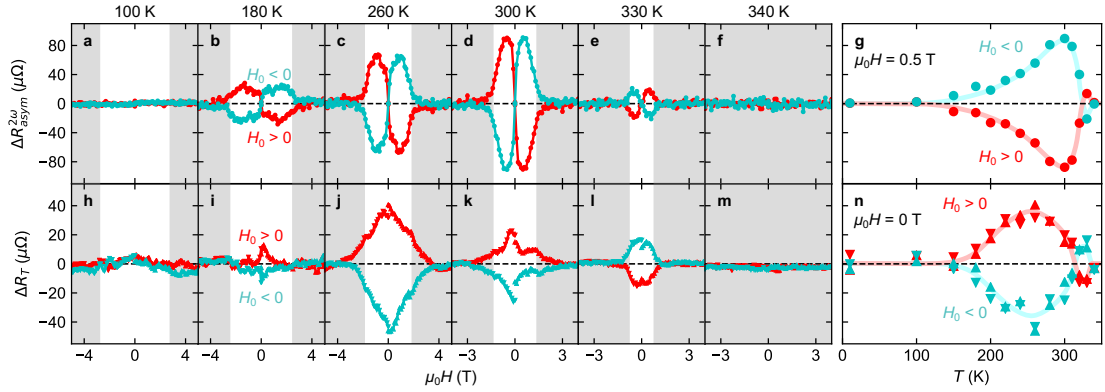

**Figure 3**

**Temperature dependence of the chirality-dependent transverse resistance.**

**a-f** Magnetic field dependence of the averaged NET signal $\Delta R^{2\omega}_{asym} = (R^{2\omega}_{asym}(+I_0) - R^{2\omega}_{asym}(-I_0))/2$ at various temperatures. Red and cyan symbols show $\Delta R^{2\omega}_{asym}$ for $H_0 > 0$ and $H_0 < 0$, respectively. While the data of $H < 0$ are merely copies of the $H > 0$ data, we plot the $H < 0$ data just for clarity. The gray shading represents the induced-FM or paramagnetic phases.

**g** Temperature, $T$, dependence of $\Delta R^{2\omega}_{asym}$ at 0.5 T for $H_0 > 0$ (red) and $H_0 < 0$ (cyan). Solid lines are guides to the eyes.

**h-m** Magnetic field dependence of $\Delta R_T$ at various temperatures. Red and cyan symbols show $\Delta R_T$ for $H_0 > 0$ and $H_0 < 0$, respectively.

**n** $T$ dependence of $\Delta R_T$ at 0 T for $H_0 > 0$ (red) and $H_0 < 0$ (cyan). Solid lines are guides to the eyes.

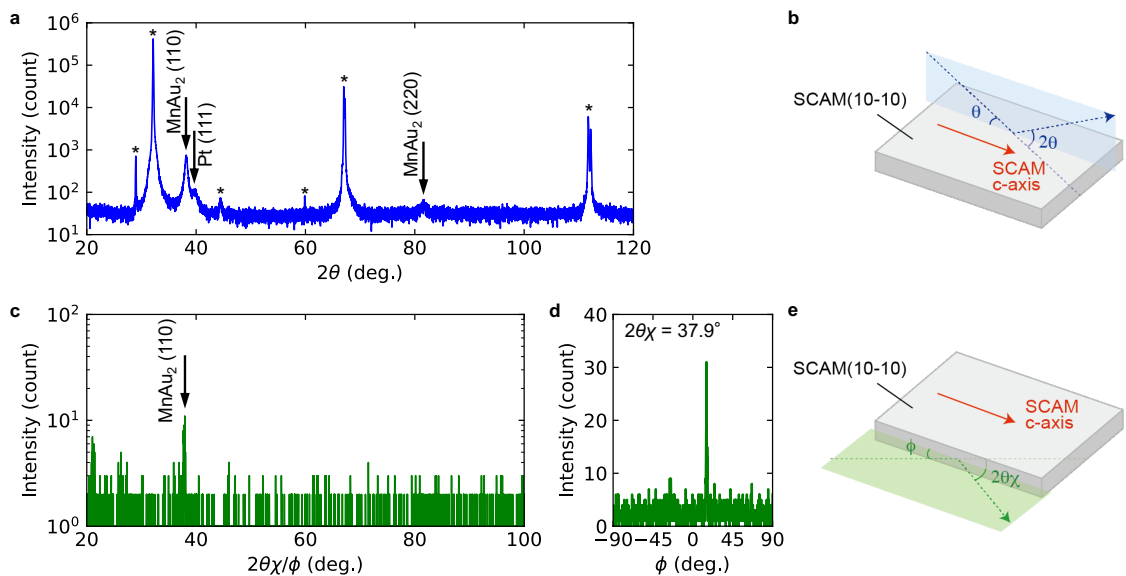

**Extended Data Figure 1**

**XRD profiles of the MnAu$_2$ / Pt bilayer film sample.**

**a** Out-of-plane $\theta/2\theta$-scan. MnAu$_2$ (110) and (220) reflections indicate that the MnAu$_2$ (110) plane is grown on the ScMgAlO$_4$ (10−10) plane. Pt (111) reflection indicates the orientation of the Pt (111) plane. Asterisks denote the peaks from the ScMgAlO$_4$ (SCAM) substrate or sample stage. No extra reflections were observed.

**b** Schematic illustration of the out-of-plane $\theta/2\theta$-scan.

**c** In-plane $2\theta\chi/\varphi$-scan. The MnAu$_2$ (110) reflection indicates that the MnAu$_2$ [1−10] direction is perpendicular to the ScMgAlO$_4$ [0001] direction.

**d** In-plane $\varphi$-scan. A single sharp peak in the $\varphi$-scan range from −90° to 90° indicates the in-plane single-crystal order of the MnAu$_2$ layer.

**e** Schematic illustration of the in-plane $2\theta\chi/\varphi$- and $\varphi$- scans.

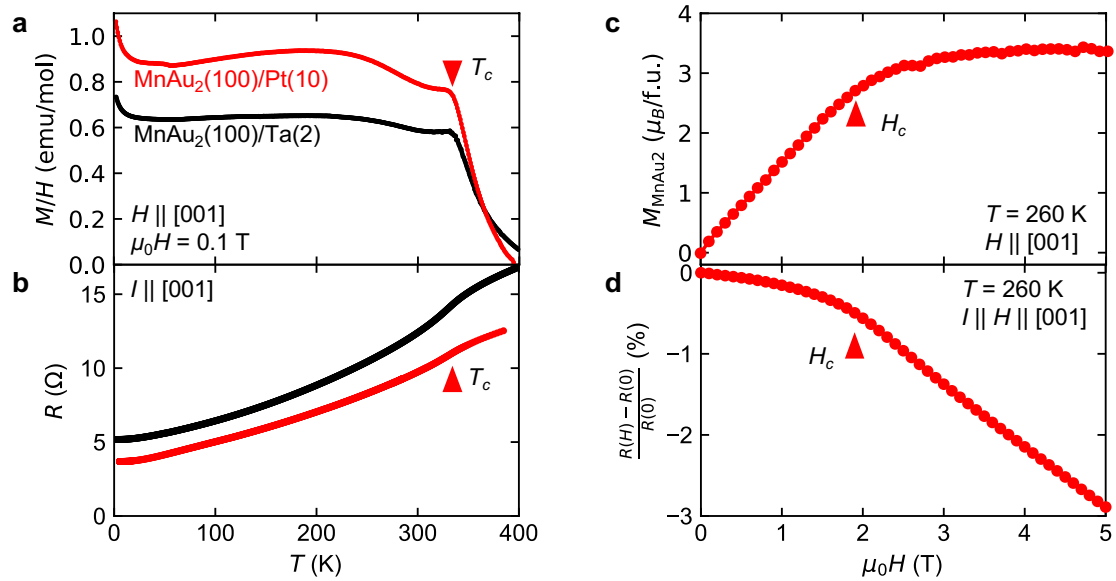

**Extended Data Figure 2**

**Properties of the MnAu$_2$ / Pt bilayer film sample.**

**a** Temperature, $T$, dependence of the magnetic susceptibility $M/H$ of the MnAu$_2$ / Pt sample, which is obtained by the magnetization $M$ divided by the magnetic field $H$ at 0.1 T. The data for the MnAu$_2$ (100nm) / Ta (2 nm) sample are reproduced from ref. 18. The triangle denotes the helimagnetic transition temperature $T_c$ = 335 K.

**b** $T$ dependence of the longitudinal resistance $R$. The data for the MnAu$_2$ (100nm) / Ta (2 nm) sample are reproduced from ref. 18.

**c** Magnetic field $H$ dependence of the magnetization $M_{\text{MnAu2}}$ at 260 K. The linear diamagnetic contribution is subtracted from the measured magnetization. The triangle denotes the helimagnetic to ferromagnetic transition field $H_c$.

**d** $H$ dependence of the longitudinal magnetoresistance $(R(H) - R(0)) / R(0)$ at 260 K. The triangle denotes the helimagnetic to ferromagnetic transition field $H_c$.

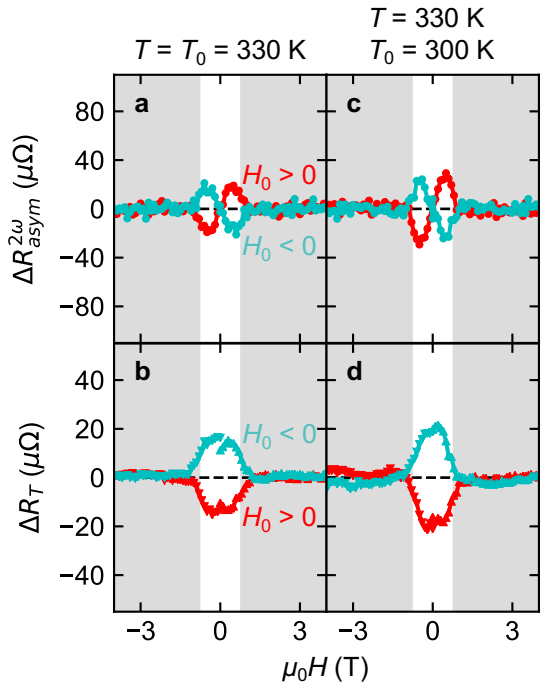

**Extended Data Figure 3**

**Comparison of nonreciprocal electrical transports and transverse resistances after chirality control at 300 K and 330 K.**

**a, b** Magnetic field $H$ dependence of $\Delta R^{2\omega}_{asym}$ (**a**) and $\Delta R_T$ (**b**) at 330 K for the chirality control temperature $T_0 = 330$ K.

**c, d** Magnetic field $H$ dependence of $\Delta R^{2\omega}_{asym}$ (**c**) and $\Delta R_T$ (**d**) at 330 K for the chirality control temperature $T_0 = 300$ K. The signs were the same as those for $T_0 = 330$ K shown in **a** and **b**, which indicates that the sign changes of $\Delta R^{2\omega}_{asym}$ and $\Delta R_T$ are not induced by the reversal of controlled chirality.